\documentstyle[12pt]{article}
 \newcommand{\be}{\begin{equation}} \newcommand{\ee}{\end{equation}}
 \newcommand{\ba}{\begin{array}{1}} \newcommand{\ea}{\end{array}}
 \newcommand{\bb}{\begin{thebibliography}}
 \newcommand{\eb}{\end{thebibliography}}
 \newcommand{\ci}[1]{\cite{#1}}
 \newcommand{\bi}[1]{\bibitem{#1}}

 \newcommand{\bk}{{\bf k}}
 \newcommand{\bp}{{\bf p}}

 \newcommand{\abstitle}[1]{\small {\bf #1}}
 \newcommand{\absauthor}[1]{\small {\bf #1}}
 \newcommand{\address}[1]{\it #1}

\begin{document}
 \setlength{\unitlength}{1mm}
 \begin{center}
 \abstitle{DELTA-ISOBAR PRODUCTION IN ANTIPROTON ANNIHILATION ON
 THE DEUTERON}\\[2.0MM]
\absauthor{G.I. Lykasov $^a$, M.P. Bussa $^b$ \footnote{Corresponding Author. 
 E-mail address: bussa@to.infn.it. Tel: +39-11-6707472. Fax: +39-11-6707476}, 
L. Valacca $^b$}\\[2.0mm]

\address{$^a$ Joint Institute for Nuclear Research, Dubna, Moscow
 Region, 141980, Russia}

 \address{$^b$ Dipartimento di Fisica, Universita' di Torino and INFN,
 Sezione di Torino, Turin, Italy.}\\

 \end{center}
 \begin{center}
 {\bf Abstract}
 \end{center}

 The annihilation of antiprotons on deuterons at rest
 via the channels  $\bar p d\rightarrow \pi \Delta$ and
 $\bar p d\rightarrow \pi N$ is studied.
 The two-step mechanism is investigated by analysing these processes
 when either $\pi^0 n$, $\pi^0 \Delta^0$
 or $\pi^- p$, $\pi^-\Delta^+$ are produced in the final state.
 Predictions
 for the branching ratios of the annihilations $\bar p d\rightarrow \pi^- p$
 and $\bar p d\rightarrow \pi^- \Delta^+$ are presented.
 From comparison of the presented results with the experimental data
 new information on the mechanism of the $\Delta$-isobar
 production in antiproton-deuteron annihilation can be deduced.
 \vspace{1cm}

 \noindent

 PACS numbers: 13.60.Le,25.30Fj,25.30Rw

 \vspace{1cm}

 It is known that antiproton annihilation on a deuteron occurs
 mostly on quasi-free nucleons leaving one nucleon a
 "spectator" with a typical momentum below 100 MeV/c. But this
 so-called impulse approximation holds until the "spectator"
 nucleon is measured. However, in the case of annihilation
 when the final baryon has a higher momentum,
 the main mechanism en\-vi\-sa\-ged is a two-step process. The antiproton
 annihilates on one nucleon producing at least two mesons, one of
 which is absorbed by the second nucleon forming
 a nucleon or a baryonic resonance, for example, a $\Delta$-isobar.
 Pontecorvo reactions such as $\bar p d\rightarrow \pi N$,
 where the final nucleon has a momentum of the order of $1$ GeV/c,
 have already been considered in the framework of this mechanism
 in refs. \ci{kondr,guar} and \ci{kudr,kutar}.
 In this paper, annihilation processes at rest on the deuteron of the
 type of $\bar p d\rightarrow \pi N$ and $\bar p d\rightarrow \pi \Delta$
 are a\-na\-ly\-sed. Some predictions for branching ratios
 and the ratio between reactions $\bar p d\rightarrow \pi^{-} p$
 and $\bar p d\rightarrow \pi^{-} \Delta^{+}$ are presented. Moreover,
 the bran\-ching ratios and the corresponding ratio between the reactions
 $\bar p d\rightarrow \pi^{0} n$ and
 $\bar p d\rightarrow \pi^{0} \Delta^{0}$ are calculated and
 compared with the experimental data of the Crystal Barrel
 Collaboration \ci{4,8} and KEK \ci{5}.
 The analysis of these processes is performed within
 the framework of the two-step mechanism of $\bar p d$ annihilation.
 Comparison of these predictions with the
 experimental data results in a new  view
 on the mechanism of the absorption process of the virtual meson
 by the virtual nucleon inside the deuteron, especially in the
 case when the $\Delta$-isobar is produced.

 The general expression for the cross section of the annihilation
 $\bar p d\rightarrow \pi B$ in flight where $B$ is a baryon, either
 a nucleon or a $\Delta$-isobar, can be written in the following
 form:
 \be
 \sigma_{\bar p d\rightarrow B}(s)=\frac{{(2\pi)}^4}{F}\int
 \delta^4(p_{\bar p}+p_d-p_{\pi}-p_B){\mid F_{\bar p d\rightarrow \pi B}
 \mid}^2\frac{d^3p_{\pi}d^3p_B}{{(2\pi)}^3 2E_{\pi}{(2\pi)}^3 2E_B}
 \hspace{0.3cm}, 
 \ee
 where $F$ is the so-called flux factor, i.e.,
 $F=2\lambda^{1/2}(s,m^2,M_d^2)=2(s-(m+M_d)^2)^{1/2}(s-(M_d-m)^2)^{1/2}$,
 here $s$ is the square initial energy in the c.m.s. of $\bar p d$,
 $m$ and $M_d$ are the nucleon and deuteron masses respectively;
 $F_{\bar p d\rightarrow \pi B}$ is the amplitude of the reaction.
 Eq.(1) holds valid for annihilation in flight
 but cannot be calculated at rest, because
 at rest the flux factor is zero. Therefore, in the case of
 annihilation at rest it would be better to analyse the branching
 ratio of the  reaction,  $Br_{\bar p d\rightarrow \pi B}=
 \sigma_{\bar p d\rightarrow \pi B}/\sigma^{tot}_{\bar p d}$,
 where $\sigma^{tot}_{\bar p d}$ is the total cross section of $\bar p d$
 annihilation.

 The amplitude $F_{\bar p d\rightarrow \pi B}$, according to the above,
 can be calculated within the framework of the two-step mechanism which in the
 general form is presented in Fig.1. It has been shown
 in refs.\ci{kudr,kutar} that
 the $\pi$-meson exchange in the triangle graph for the Pontecorvo reaction
 (see Fig.1) $\bar p d\rightarrow \pi N$
 yields the dominant contribution, about 90\% as compared with the
 $\rho,\omega$ contribution.
 One can suggest the same for $\bar p d\rightarrow \pi \Delta$
 annihilation, too, for two reasons: first,  
 the excitation of $\Delta$ by $\rho$ and $\omega$
 absorption has a smaller pro\-ba\-bi\-li\-ty than the simple absorption
 of the virtual pion by the nucleon because the off-shellness of $\rho$- or
 $\omega$- mesons is very large in comparison to the corresponding one
 of the pion; second, $\Delta$(1232) decays mainly into a pion and a nucleon.
 Therefore, one may take only the $\pi$-meson exchange into account and
 calculate the triangle graph presented in Figs.(1,2).



 The general form of the amplitude $F_{\bar p d\rightarrow \pi B}$
 corresponding to the graphs of Figs.(1,2) can be written, for example,
 within the framework of the formalism presented in \ci{dol,lyk},
 as follows:
 \be
 F_{\bar p d\rightarrow \pi B}=-\frac{1}{(2\pi)^3}\int
 \Phi_d(k)F^B_{\pi} \Gamma_{\pi N B} G^B_{\pi}(\bk)f_{\bar p N\rightarrow
 \pi \pi}\frac{d^3k}{4E(\bk)\epsilon_{\pi}({\bf q}_\pi)}
 \hspace{0.3cm} ,
 \label{gl}
 \ee
 here $\Phi_d(k)$ is the deuteron wave function (d.w.f.),
 $f_{\bar p N\rightarrow \pi \pi}$ is the amplitude of the annihilation
 process $\bar p N\rightarrow \pi \pi$, $F^B_\pi$ is the formfactor 
 corresponding to the vertex $\pi N B$,
 $G^B_\pi(\bk)$ is the propagator of the virtual $\pi$-meson which has the
 following form:
 \be
 G^B_\pi(\bk)=(E_B(\bp _B)-E(\bk)-\epsilon_{\pi}({\bf q}_\pi)+i\epsilon)^{-1}
 \hspace{0.3cm} ,
 \ee
 where $E_B({\bf p}_B)$ and $E({\bf k})$ are the energies of the final
 baryon and of the internal nucleon inside the deuteron;
 $\epsilon_{\pi}({\bf q}_{\pi})$ is the energy of the virtual
 $\pi$-meson in the intermediate state,
 ${\bf q}_{\pi}={\bf p}_B-\bf k$
 is its three-momentum; ${\bf p}_B$ is the three-momentum of the final
 baryon; $\Gamma_{\pi N B}$ corresponds to the lower vertex
 of Fig.1 or Fig.2, i.e. to absorption of the virtual pion by the
 nucleon inside the deuteron, in particular,
 $\Gamma_{\pi N N} = g_{\pi N N }$ ({\boldmath $ \sigma \cdot \tau$}),
 {\boldmath $\tau$}  = $ b(\frac{\bf p_N}{E_N(\bp_3)+m} -
 \frac{\bf k}{E(\bk)+m})$,
 $b = (E(\bk) + m)(E_N(\bp_N) + m)/2m $; {\boldmath $ \sigma$} is the vector 
 Pauli matrix, $g_{\pi N N}$ is the corresponding coupling constant,
 $g^2_{\pi NN}/4\pi$=14.7 \ci{1,2}; $\Gamma_{\pi N \Delta} = g_{\pi N\Delta}
 (E(\bk) + m)^{1/2}(E_\Delta (\bp_\Delta) + m_\Delta)^{1/2}$, $g_{\pi N\Delta}$
 is the $\pi N\Delta$ coupling constant and
 $g^2_{\pi N\Delta}/4\pi$ = 56$\div$71 according to \ci{1,2} and \ci{hol,kisl}.

 Substituting $F_{\bar p d\rightarrow \pi B}$  into eq.(1) and
 neglecting the interference between different isotopic triangle
 graphs we can get the branching ratios of the processes
 $\bar p d\rightarrow \pi N$ and $\bar p d\rightarrow \pi \Delta$.
 In particular, for the final states containing the neutral particles
 $\pi^0 n$ and $\pi^0 \Delta^0$ we have:
 \be
 Br_{\bar p d\rightarrow \pi^0 n} = (2 Br_{\bar p n\rightarrow
 \pi^- \pi^0} + Br_{\bar p p\rightarrow \pi^0 \pi^0}) c_N
 \mid J_N\mid^2g^2_{\pi NN}
 \hspace{0.3cm} ,
 \label{10}
 \ee
 \be
 Br_{\bar p d\rightarrow \pi^0 \Delta^0} = (2
 Br_{\bar p n\rightarrow \pi^- \pi^0} + Br_{\bar p p\rightarrow
 \pi^0 \pi^0}) c_\Delta
 \mid J_{\Delta}\mid^2
 g^2_{\pi \Delta N}
 \hspace{0.3cm} ,
 \label{11}
 \ee
 where $Br_{\bar p n\rightarrow \pi^-\pi^0}$ and
 $Br_{\bar p p\rightarrow \pi^0\pi^0}$
 are the branching ratios of the annihilation processes
 $\bar p n\rightarrow \pi^-\pi^0$
 and $\bar p p\rightarrow \pi^0 \pi^0$ at rest, respectively,
 $c_N = 1/E_N(\bp _N)$ and
 $c_\Delta = (E_\Delta(\bp_\Delta) + m)/E_\Delta(\bp_\Delta)$.
 Eqs. (\ref{10}-\ref{11}) were derived under
 the assumption that d.w.f. decreases with
 the internal momentum $k$ more rapidly than the amplitude
 $f_{\bar p N\rightarrow \pi \pi}$. The expressions
 for the integrals $J_N$ and $J_\Delta$ are the following:
 \be
 J_N=\frac{1}{4\sqrt{2\pi}}\int_{-1}^1 d\cos\theta(k)_{p}\int_0^{k_{max}}
 \Phi_d(k)
 \tau_z F^N_\pi (q^2)G_N(\bk)\frac{k^2dk}{E(\bk)}
 \label{g5}
 \ee
 and
 \be
 J_\Delta=\frac{1}{4\sqrt{2\pi}}\int_{-1}^1d\cos\theta(k)_{\Delta}
\int_0^{k_{max}}\Phi_d(k)
 (E(\bk)+m)^{1/2}F^{\Delta}_\pi (q^2)G_\Delta (\bk)\frac{k^2dk}{E(\bk)}
 \hspace{0.3cm} ,
 \label{g6}
 \ee
 where $k_{max}\simeq 0.8-1.0$ GeV/c is the maximum value of the internal 
 momentum of the nucleon inside of the deuteron,  
 $F^N_\pi$ and $F^{\Delta}_\pi$ are the
 pion formfactors for the processes $\bar p d\rightarrow \pi N$ and
 $\bar p d\rightarrow \pi \Delta$, respectively,
 $\tau_z = b({\bf p}_N/(E_N(\bp_N) + m) - {\bf k}/(E(\bk) + m)){\bf p}_N
 /\mid {\bf p}_N\mid$; $\theta(k)_{N}$ and $\theta(k)_{\Delta}$ are the 
 angles between the momenta of the final baryon, $N$ or $\Delta$, and 
 the internal nucleon inside of the deuteron. 
 
 The ratio $R_1$ between
 (\ref{10}) and (\ref{11}) has the following form:
 \be
 R_1 = \frac{Br_{\bar p d\rightarrow \pi^0 n }}{Br_{\bar p d\rightarrow \pi^0 \Delta^0}}
 \simeq \frac{\mid J_N\mid^2}{\mid J_\Delta\mid^2}
 \frac{g^2_{\pi NN}}{g^2_{\pi \Delta N}}\frac{c_N}{c_\Delta}
 \hspace{0.3cm} .
 \label{8}
 \ee
 Under isospin invariance for the  $|\pi^0 n>$,
 $|\pi^0 \Delta^0>$ and $|\pi^- p>$, $|\pi^- \Delta^+>$ final states
 the corresponding relations can be deduced for $\bar p d\rightarrow \pi^- p$
 and $\bar p d\rightarrow \pi^- \Delta^+$ annihilation processes:
 \be
 Br_{\bar p d\rightarrow \pi^- p} = 2Br_{\bar p d\rightarrow \pi^0 n}
 \label{6}
 \ee
 \be
 Br_{\bar p d\rightarrow \pi^- \Delta^+} = \frac{1}{2}
 Br_{\bar p d\rightarrow \pi^0 \Delta^0}
 \label{7}
 \ee
 The corresponding expected ratio of the branchings is:
 \be
 R_2 = \frac{Br_{\bar p d\rightarrow \pi^- p}}
 {Br_{\bar p d\rightarrow \pi^- \Delta^+}} = 4R_1
 \hspace{0.3cm} .
 \label{9}
 \ee

 \newpage

 $\bullet~~$ {\bf  Kinematics}

 The conservation law at the vertex of the absorption of the
 virtual $\pi$-meson by the nucleon inside the deuteron  is
 expressed by
 \be
 (q+k)^2=m^2_B
 \label{13}
 \ee
 where $q$ and $k$ are the four-momenta of the $\pi$-meson and 
 the above mentioned nucleon, respectively; $m_B$ is the mass of either the
 nucleon or the $\Delta$-isobar.

 Solution of eq.(\ref{13}) yields the following form 
 for the square four-momentum of the virtual pion:
 \be
 q^2=m_B^2 + m^2 + 2({\bf p}_B{\bf k} - EE_B)
 \hspace{0.3cm} ,
 \ee
 where ${\bf p}_B$ is the three-momentum of the outgoing baryon $B$, i.e.
 of either the nucleon or the $\Delta$-isobar.

 The momentum of particle $B$ can be deduced from the conservation
 law for the Pontecorvo reaction at rest:
 \be
 (p_\pi+p_B)^2=(m_N+m_d)^2
 \hspace{0.3cm} ,
 \ee
 here $p_\pi$ and $p_B$ are the four-momenta of the final $\pi$-meson
 and nucleon or $\Delta$, respectively; $m_N$ and $m_d$ are the nucleon and
 deuteron masses, respectively. Finally, the square
 three-momenta of the final nucleon or $\Delta$-isobar are respectively:
 \be
 |\bp _N|^2=\frac{a_N^2-\mu^2 m_N^2}{9m_N^2}
 \ee
 and
 \be
 |\bp _\Delta|^2=\frac{a_\Delta^2-\mu^2 m_\Delta^2}{9m_N^2}
 \hspace{0.3cm} ,
 \ee
 where for the final nucleon $a_N\simeq 4m_N^2-\mu^2/2$ and for the
 $\Delta$-isobar $a_\Delta\simeq \frac{1}{2}(9m_N^2-m_\Delta^2-\mu^2)$,
 here the binding energy of the deuteron ($\simeq 2.2$ MeV) is neglected
 and $(m_N+m_d)^2\simeq 9m_N^2$ is assumed.

 $\bullet~~$ {\bf  Formfactors \boldmath $F^N_\pi (q^2)$ and
  \boldmath $F^\Delta_\pi (q^2)$}

 The choice of the formfactors (FF) $F^N_\pi (q^2)$
 and $F^\Delta_\pi (q^2)$
 in eqs. (\ref{g5}) and (\ref{g6}) is very important.
 For example, for the annihilation $\bar pd\rightarrow \pi B$
 the form factor can be chosen in the following form \ci{2,hol}
 \be
 F_\pi^B (q^2)=(\frac{\Lambda_B^2 - \mu^2}{\Lambda_B^2 + |q^2|})^n
 \label{l1}
 \ee
 where either $n=1$ (so called monopole FF) or $n=2$ (dipole FF).

 In principle, the values of the cut-off parameter $\Lambda_B$
 and of $n$ could be different
 for $\bar p d\rightarrow \pi N$ and $\bar p d\rightarrow
 \pi \Delta$ annihilations.
 For example, according to refs.\ci{kondr,guar} and \ci{kudr},
 $\Lambda_N=1.2-1.4$ GeV/c and $n=1$ for the first process.
 As to the second reaction, the form of the FF
 can be chosen like in the description of the processes
 $\Delta N\rightarrow \Delta N, \Delta N\rightarrow \Delta \Delta$ etc.
 For example, following \ci{kisl} we can use 
 \be
 \label{l2}
  F_{\Delta} (q^2)=(\frac{\Lambda_{\Delta}^2}{\Lambda_{\Delta}^2 + |q^2|})^n
  \ee
 or, according to \ci{7}, the exponential form:
 \be
 \label{exp}
  F_{\Delta} (q^2)=exp(-{\bf q}^2/(2\Lambda_{\Delta}^2)
 \ee
 We calculated the branchings using these FF (\ref{l2}) and (\ref{exp}).
 However the exponential form of FF (\ref{exp}) didn't reproduce the 
 experimental data by applying the values of cut-off parameter 
 $\Lambda_{\Delta}$ = (2 - 3)$\mu$ presented by \ci{7}. Therefore 
 we limited ourselves to the application of the more conventional FF
 of type (\ref{l2}).

 $\bullet~~$ {\bf  Results and Discussions.}

 We now present the resultant calculated branching ratios
 $BR$ for the reactions $\bar p d\rightarrow \pi N$ and
 $\bar p d\rightarrow \pi \Delta$.  They were calculated
 using different forms for the FF and different values of
 the cut-off parameters $\Lambda_N$ and $\Lambda_{\Delta }$.\\
 The results obtained for the annihilation $\bar pd\rightarrow n \pi^0$
 using Paris d.w.f. \ci{lacomb} and FF
 (\ref{l1}) with the $n=1$ standard choice are presented in Table 1 for
 different values of the cut-off parameter $\Lambda_N$.
 As a consequence of the monopole choice, the range of values of the
 cut-off parameter to be explored should be close to the nucleon mass
 (see, for example, \ci{OTW} and references quoted therein), 
 as confirmed also by the results presented in
 \ci{kondr}, \ci{guar} and \ci{kudr}.\
 Using FF (\ref{l2}) with $n=2$ for
 the annihilation $\bar pd\rightarrow \pi^0 \Delta^0$ and the
 coupling constant $g^2_{\pi N\Delta}/4\pi$=71 according to \ci{hol,12}
 we obtain the plot of the branching ratio versus $\Lambda_\Delta$
 presented in Fig.3. When FF (\ref{l2}) is adopted 
 - or, equivalently, FF (\ref{l1}) -
 with the monopole choice it is impossible to reproduce the experimental data
 by keeping the cut-off parameter reasonably close to the nucleon mass. 
 \begin{table}[htb]
 \begin{center}
 \caption{BR for Pontecorvo reaction: $\bar p+d\rightarrow n+\pi^0$.}
 \vspace*{.5 cm}
 \begin{tabular}{|c|c|}
 \hline
 $\Lambda_{N}$(GeV/c) & BR ($\times 10^{-6}$)\\
 \hline
 1.1 &  6.278 \\
 1.2 &  7.432 \\
 1.3 &  8.544 \\
 1.4 &  9.600 \\
 \hline
 \end{tabular}

 \end{center}
 \end{table}

 Note that the presented branchings and ratios
 reproduce the ex\-pe\-ri\-men\-tal data of Crystal Barrel Collaboration : 
 $BR(\bar p d\rightarrow \Delta^0(1232) \pi^0)=
 (22.1 \pm 2.4)10^{-6}$,
 $BR(\bar p d\rightarrow n \pi^0)=(7.3 \pm 0.72)10^{-6}$ 
 and $R_1=0.32 \pm 0.05$ \ci{4,8}.


 The corresponding branching ratios for the reactions
 $\bar p d\rightarrow \pi^- p$ and 
 $\bar p d\rightarrow \pi^-\Delta^+$ 
 can be obtained from Table 1  and Fig.3 by multiplying
 the branching ratios for $\bar p d\rightarrow n \pi^0$
 by a factor 2 and for $\bar p d\rightarrow \pi^0\Delta^0$
 by a factor 1/2, in accordance with the isotopic relations 
 (\ref{6}) and (\ref{7}) respectively. 

 $\bullet~~$ {\bf Conclusions.}

 The two-step approach applied to Pontecorvo reactions $\bar p d\rightarrow \pi
 N$ results in branching ratios for $\bar p d\rightarrow \pi^0 n$
 annihilation which do not contradict exi\-sting experimental data \ci{8}
 within the experimental errors, quite large in the case of the KEK data
 \ci{5}.    
 A good agreement is also observed with the calculated BR and the most recent
 measurement on the reaction $\bar p d\rightarrow \pi^- p$,
 performed at the OBELIX Spectrometer \ci{obe}.\\
 Our approach is close to the one considered in ref.\ci{kutar} and the 
 results obtained in this paper for $BR(\bar p d\rightarrow \pi^0 n)$
 approximately coincide with the corresponding ones presented there.
 This result may serve as a basis for the application 
 of our approach to pion and
 $\Delta$-isobar production at rest. The triangle graph for
 $\bar p d\rightarrow \pi\Delta$ is distinguished from the one
 for $\bar p d\rightarrow \pi N$
 annihilation by the different values of coupling constants
 ($g^2_{\pi N\Delta}/g^2_{\pi N N}\simeq 4-5$) and by different spin structures
 of the corresponding pion-nucleon absorption vertices. Moreover, the
 pion propagators for the corresponding two graphs (Fig.(1,2)) 
 have dif\-fe\-rent pole
 positions in the calculation of eqs. (\ref{g5},\ref{g6}).
 Such a difference between calculations of the triangle graphs
 for $\bar p d\rightarrow \pi N$ and
 $\bar p d\rightarrow \pi\Delta$ annihilations allows one to assume that
 the forms of the FF should also be different in order to describe the
 experimental data on the branching ratio for
 $\bar p d\rightarrow \pi^0 \Delta^0$ \ci{4}.
 By using a Paris d.w.f. the best fit of CB experimental data
 for $\bar p d\rightarrow \pi^0 n$ annihilation is obtained with monopole
 FF (\ref{l1}) and cut-off parameter $\Lambda_N=1.1-1.2$ GeV/c.
 In order to describe the CB measurements on the branching ratio for 
 the reaction $\bar p d\rightarrow \pi^0 \Delta ^0$, with quite the same
 value of the previous cut-off parameter ($\Lambda_{\Delta}\simeq 1.$ GeV/c),
 a FF (\ref{l1}) of a form of higher order than the monopole is required: 
 that is the dipole formfactor. 
 Using this value of $\Lambda_\Delta$, the results presented in Fig.3 and the
 isotopic relation (\ref{7}) one can show the branching ratio for 
 $\bar p d\rightarrow \pi^-\Delta^+$ annihilation to be about $(10-12)*10{^-6}$.
 This result can be considered as a certain prediction for the OBELIX
 experiment, that is analysing this channel \ci{CIPANP97}. 
 The application of other forms of FF doesn't result
 in a satisfactory description of the experimental data.

 This does not contradict the previous study of Pontecorvo
 reactions \ci{kondr,guar,kudr,kutar} and means that
 annihilation of the type of $\bar p d\rightarrow \pi \Delta$ can
 be described with the help of the two-step mechanism without 
 introduction of any non-nucleonic six-quark component
 in the deuteron successfully applied in the description of the prototype
 Pontecorvo reaction \ci{guar} and recently invoked to describe
 the baryon-baryon content of the deuteron \ci{10}. 
 Nevetherless, in the framework of
 the two-step model, the investigation of $\Delta$-
 isobar production in antiproton-deuteron annihilation
 results in a new view on its production mechanism and, in
 particular, in new information on the choice of the formfactor
 of the virtual pion and of the cut-off parameter $\Lambda_{\Delta}$
 which is still quite insufficient \ci{2,hol}.         \\
 We gratefully acknowledge the very helpful discussions with 
 S.Moszkovsky, R.Machleidt and O.Denisov. We are grateful to
 G.Pontecorvo for the help in preparing this paper.

 

 \end{document}